\documentclass[useAMS,usenatbib]{mn2e}
\usepackage{graphicx}
\usepackage{epsfig}
\def\lesssim{\la}
\def\gtrsim{\ga}

\title[]{Three Body Resonance Overlap in Closely Spaced Multiple Planet Systems}
\author[]{Alice C. Quillen  \\
{Department of Physics and Astronomy, University of Rochester, Rochester, NY 14627, USA}  
}

\begin{document}
\label{firstpage}
\maketitle

\begin{abstract}

We compute the strengths of zero-th order (in eccentricity) three-body resonances for  a 
co-planar and low eccentricity multiple planet system.  
In a numerical integration we illustrate that slowly moving Laplace angles
are matched by variations in semi-major axes among three bodies with the outer two bodies moving
in the same direction and the inner one moving in the opposite direction, as would be expected from the two 
quantities that are conserved in the three-body resonance.      
A resonance overlap criterion is derived for the closely and uniformly spaced, equal mass system 
with three-body resonances overlapping when interplanetary
separation is less than an order unity factor times the planet mass to the one quarter power.      
We find that three-body resonances are sufficiently
dense to account for wander in semi-major axis seen in numerical integrations of
closely spaced systems and they are likely the cause of
instability of these systems.   For interplanetary separations  
outside the overlap region, stability timescales significantly increase.
Crudely estimated diffusion coefficients in eccentricity and semi-major axis
depend on a high power of planet mass and interplanetary spacing.   
An exponential dependence previously fit to stability or crossing timescales is likely due to 
the limited range of parameters and times possible in integration and the strong power law dependence
of the diffusion rates on these quantities.

%numerically is due to the steep dependence of libration timescales in these resonances on the planetary separation.
%and the timescale it takes for semi-major axis wander to encounter second order resonances?  
\end{abstract}

\section{Introduction}

The stability of multiple planet systems has been long been a matter
of interest as it concerns the long term stability of the Solar system.
Poincar\'e noticed that perturbation techniques involved singularities or 
small divisors that prevented solution by convergent series. 
Recent numerical explorations suggest that
the giant planets in our solar system were originally located in a more
compact location and experienced a subsequent planet-planet scattering event 
(within the context of the ``Nice model''; \citealt{tsiganis05}).
All extrasolar planetary systems may experience epochs of dynamically instability 
(e.g. \citealt{ford01,barnes06,chatter08,thommes08,goz08,goz09,raymond09a,kopparapu10,fabrycky10}).

Numerical integrations show that
a system of two planets on initially zero-inclination and
eccentricity orbits about a star never experience mutual close encounters 
if the initial semimajor axis separation is sufficiently large 
\citep{marchal82,gladman93,barnes07}, (also see \citealt{mardling08}). 
%$\Delta_m$, measured in the mutual 
%Hill radii, exceeds $2\sqrt{3}$ \citep{gladman93}.
Stability of multiple planet systems is often discussed in terms
of this limit which has been called ``Hill stability'' 
(e.g., \citealt{marchal82,barnes07,raymond09b}).
Systems with multiple planets or satellites are stable over long periods
of time if the bodies are sufficiently distant from each other 
\citep{chambers96,duncan97,faber07,chatter08,smith09}.
For an integration begun with all bodies at zero eccentricity and inclination, the integration time until one
body crosses the orbit of an other body is known as a crossing timescale, $t_c$.
We refer to the idealized problem studied by \citet{chambers96} with equal mass planets and 
with the semi-major axis of each consecutive planet set
from that of the previous one using a constant $\delta$,
\begin{equation}
a_{n+1} = (1+ \delta) a_n.
\label{eqn:delta}  
\end{equation}
The planets have mass ratio with respect to the central star $m = m_p/M_*$.   
When the planet number is greater than five, the crossing timescale is insensitive 
to the number of bodies (e.g., \citealt{chambers96}).

A number of studies have fit power laws to the crossing or stability timescale.
\citet{duncan97} found that the crossing time is sensitive to the mass of the satellites with $\log t_c \sim \alpha m + \beta$ where $\alpha$ is a slope and $\beta$ is an offset.
\citet{chambers96} found  $\log t_c \sim \alpha (\delta m^{-1/4}) + \beta$, whereas
\citet{smith09} fit
$\log t_c  \sim  \alpha  (\delta m^{-1/3}) + \beta $.  Here the parameters $\alpha$, $\beta$ 
are fit to the numerically measured crossing timescales and are 
not identical in each setting.  These numerical studies integrated for timescales between 100 and $10^9$ orbital periods of the innermost orbiting body.
The interplanetary separation ranged from 1 to about 10 mutual Hill radii and  
the mass ratio ranged from $m\sim10^{-3}$ to $10^{-9}$.   
%The Uranian satellite system I think has $\delta$ of 10-14 Hill radii (Rob French, private communication).
The trends in the numerically measured crossing timescales are currently lacking an explanation.

The power law forms for the crossing timescales 
can be re-written $t_c \propto \exp ( m^\alpha \delta^\gamma)$  for exponents
$\alpha,\gamma$ \citep{faber07}.
The exponential form is reminiscent of
the Nekhoroshev theorem \citep{nek77} or of Arnold diffusion \citep{arnold62} in the context of weakly perturbed
Hamiltonian systems.
There are subtle and deep connections between the Nekhoroshev theorem and Arnold diffusion (e.g, as explored and discussed by  \citealt{chirikov79,lochak92,lochak93}).
Arnold diffusion takes place on exponentially long timescales
whereas systems with resonances that are fully overlapped can have relatively faster diffusion
rates.  This led \citet{morbi95} to suggest that systems with sparse and weak resonances might exhibit 
diffusion on  exponentially long timescales
whereas those affected by multiple and overlapping resonances would diffuse at a rate that depends on 
a power law of time (also see \citealt{guzzo02}). A number of studies have numerically measured a power law relation between Lyapunov and instability or crossing timescales \citep{lecar92,levison93,murison94,morbi95,mikkola07,urminsky09,shev10} suggesting that the stability or crossing timescales are driven by chaotic diffusion or Hamiltonian intermittency.   
Because there is a power law relation between Lyapunov and crossing timescales, we would expect that 
the diffusion rate would be a power law function of the perturbation parameters (such as planet mass and 
planetary separation) rather than an exponential function of them as commonly fit. 
%(e.g., \citep{chambers96,duncan97,zhou07,faber07,smith09}).  
This apparent contradiction has been discussed in terms of two regimes for the dynamics, a weakly perturbed 
``Nekhoroshev'' regime exhibiting diffusion over exponentially long timescales and a resonant overlap power law
 regime exhibiting diffusion at a rate that depends on a power of the perturbations \citep{morbi95}.

In this study we consider the role of three-body resonances in the idealized setting studied by \citet{chambers96} 
of a closely and uniformly spaced equal mass co-planar system initially in nearly circular orbits.   
Our goal is to understand the dynamics of the idealized coplanar, closely and uniformly spaced equal mass multiple planet system sufficiently well that 
we can identify the source of instabilities in multiple planet systems.   
Previous studies of three-body
resonances have primarily focused on settings where one of the bodies is small, such as an asteroid
perturbed by both Jupiter and Saturn \citep{murray98, nesvorny99,guzzo05} but also include the early study of
the Laplace resonance by \citet{aksnes88}.   While three-body resonances are weak there are more of
them than two-body resonances so they can be a source of chaotic behavior causing slow diffusion
in eccentricity and inclination \citep{nesvorny98,nesvorny99,guzzo02,guzzo05}.  Three body resonances
may be important in extra solar multiple planet systems.  
The long-term stability of extra solar multiple planet systems may be influenced by a net of low order two and three-body resonances \citep{goz08,goz09,fabrycky10}.

%The trends in the numerically measured crossing timescales are currently lacking an explanation.
We first estimate the strength and libration frequency in zero-th order (in eccentricity) three-body resonances. 
Conserved quantities for them are also estimated so that their signature in numerical integrations can be identified.  
Using estimated numbers and widths of the three-body resonances we derive a resonance overlap criterion.
In the final section we discuss the role of three-body resonances in causing instability
in multiple planet systems and whether they may eventually provide an
explanation for the exponential forms fit to their crossing timescales.
%A summary and discussion follows.

\section{Hamiltonian for a Multiple Planet system}

\subsection{Non-interacting System}

The Hamiltonian for $N$ non-interacting massive bodies orbiting a star (and so feeling gravity only from the star)
can be written
\begin{equation}
H_0 =  \sum_{j=1}^N - { m_j^3 \over 2 \Lambda_j^2}
\label{eqn:H0simp}
\end{equation}
where $m_j$ is the mass of the $j$-th body (or planet) divided by
the mass of the star, $M_*$. 
Here we have ignored the motion of the star and have put the above Hamiltonian 
in units such that $G M_*=1$ where $G$ is the gravitational constant.  
Here the Poincar\'e coordinate $\Lambda_j = m_j \sqrt{a_j}$,
where the semi-major axis of the $j$-th planet is $a_j$.
This Poincar\'e coordinate is conjugate to the mean longitude, $\lambda_j$ of the $j$-th body.
The mean longitude $\lambda_j = M_j + \varpi_j$ where $M_j$ is the mean anomaly 
and $\varpi_i$ is the longitude of pericenter of the $j$-th body.
We may also use the Poincar\'e coordinate 
$\Gamma_j = m_j \sqrt{a_j }(1 -\sqrt{1- e_j^2}) \approx m_j \sqrt{a_j} e_j^2/2$
where $e_j$ is the $j$-th body's eccentricity.  This coordinate
is conjugate to the angle $\gamma_j = -\varpi_j$. %where $\varpi_j$ is the longitude of pericenter.  
We will restrict our system so that all planets 
are orbiting in the same plane and so will begin by ignoring the Poincar\'e coordinates
associated with inclination and the longitude of the ascending node.
Each Poincar\'e momenta contains a factor of a planet's mass.    Some studies
of three-body resonances have focused on the problem of a low mass object in the presence of two planets 
(e.g., an asteroid perturbed by Jupiter and Saturn; \citealt{nesvorny98,murray98}) and so have removed the
mass from the Poincar\'e momenta associated with the low mass object.

%We expand the above Hamiltonian about a vector of mean motion values 
%${\bf n}_{0}$ where each component is the mean  motion of a planet. 
%It is convenient to define  \begin{equation} \Lambda_{0,j} \equiv m_j n_{0,j}^{-1/3} \end{equation}
%Expanding the unperturbed Hamiltonian
%\begin{eqnarray}
%H_0 = \sum_j \left[   n_{0,j} (\Lambda_j - \Lambda_{0,j})  
%                -  { 3 \over 2 m_j a_{0,j}^2} (\Lambda_j - \Lambda_{0,j})^2   \right] \\
%   \qquad + {\rm constant} + O(\Lambda_j^3)   \end{eqnarray}
%Let $L_j = \Lambda_j - \Lambda_{0,j}$.  To  second order in ${\bf L}$
%\begin{equation}
%H_0 = \sum_j \left[ n_{0,j} L_j - {3\over 2 a_{0,j}^2 m_j} L_j^2 \right]  + {\rm constant}
%\label{eqn:H0} \end{equation}
%The transformation is canonical and $L_j$ is the momentum conjugate to
 %the angle $\lambda_j$. Here $a_{0,j} = n_{0,j}^{-2/3}$. 

\subsection{Interactions}

\begin{figure}%[htbp]
\begin{center}
\includegraphics[width=8cm]{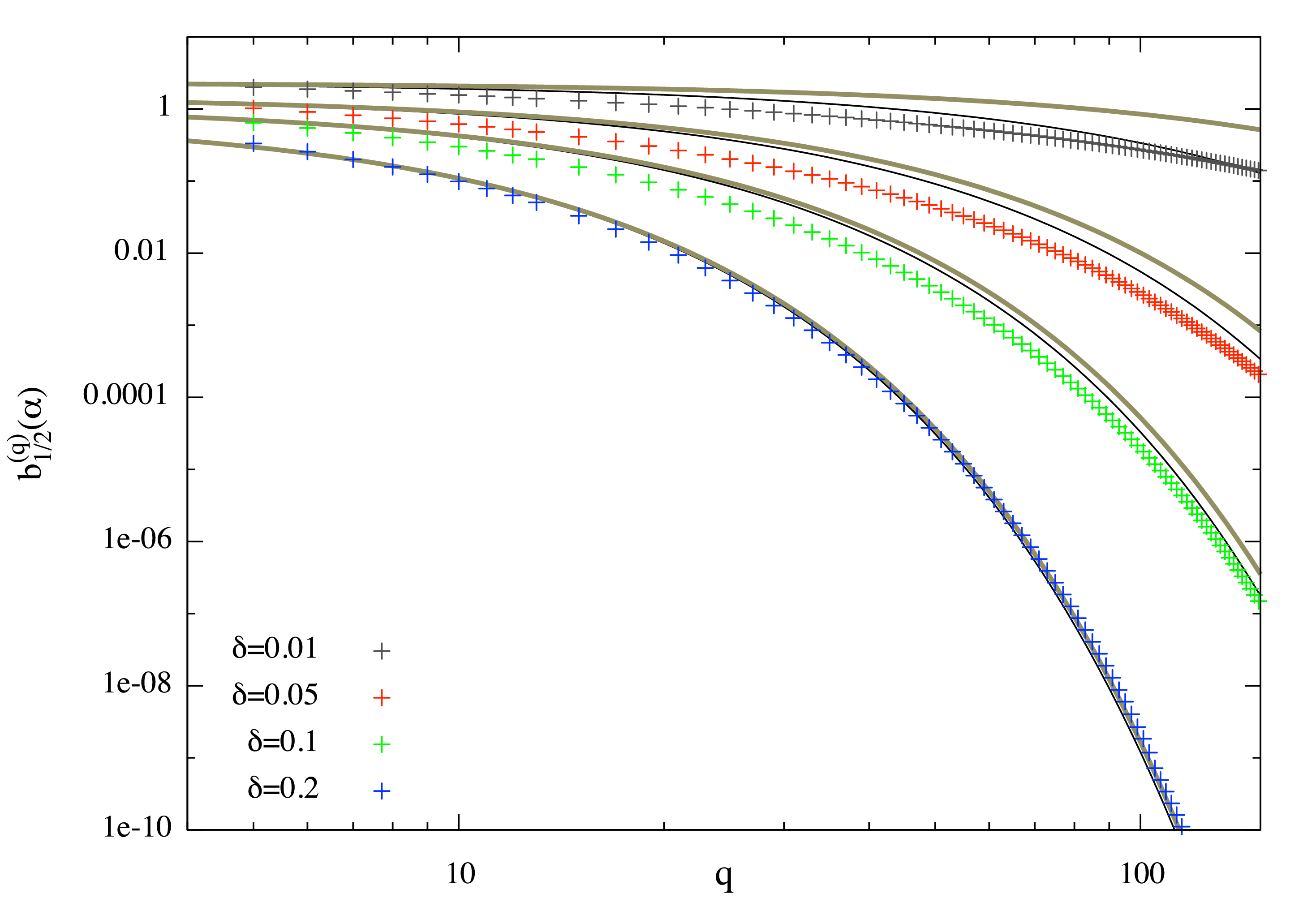}
\caption{Approximations to the Laplace coefficient.  
Plotted as points are the Laplace coefficient $b_{1/2}^{(q)}(\alpha)$ as a function of the integer $q$ 
for four separations $\delta=\alpha^{-1}-1 = 0.01,0.05,0.1,$ and $0.2$.  Overplotted as thick brown lines
for each $\delta$ value is the function $0.5 |\ln(\delta)| \exp(- q  \delta)$ and as thin black lines
$0.5 |\ln(\delta)| \exp[- q  \delta(1 -0.002 \ln \delta)]$. 
 }
\label{fig:lap}
\end{center}
\end{figure}

We now consider the gravitational interactions between the planets. 
Here we consider only the direct term and ignore the indirect terms.
The Hamiltonian can be written
\begin{equation}
H = H_0 + H_{Int}
\end{equation}
where the interaction term, $H_{Int}$, is a sum of
direct interaction terms, $H_{Int} = \sum_{j > i} W_{ij}({\bf r}_i, {\bf r}_j)$, and
\begin{equation}
W_{ij} = -{m_i m_j \over \left| {\bf r}_i - {\bf r}_j \right| } 
\label{eqn:Wij}
\end{equation}
If the planets are in nearly circular orbits
\begin{equation}
W_{ij} = -{ m_i m_j \over a_j \sqrt{1 + \alpha_{ij}^2 - 2\alpha_{ij} \cos (\lambda_i- \lambda_j)} }.
\end{equation} 
and we have assumed that $a_j > a_i$ and 
\begin{equation} \alpha_{ij} \equiv {a_i  \over a_j}.\end{equation} 
We can expand in Fourier components
\begin{equation}
W_{ij} = \sum_{q=0}^\infty  W_{ij,q} \cos( q\lambda_i - q\lambda_j)
\label{eqn:Wijf}
\end{equation}
with coefficients
\begin{equation}
W_{ij,q} =- { m_i m_j \over  a_j } b_{1/2}^{(q)} (\alpha_{ij}) 
\end{equation} 
where $b_{1/2}^{(q)}(\alpha)$ is a Laplace coefficient, 
\begin{equation}
b_{s}^{(q)}(\alpha) \equiv {1 \over \pi} \int_0^{2\pi} {\cos (q \phi) d \phi \over (1 + \alpha^2 - 2 \alpha \cos \phi )^{s}}.
\end{equation}
Laplace coefficients are the Fourier coefficients of twice the function 
$ f(\phi) = (1 + \alpha^2 - 2 \alpha \cos \phi)^{-s}$.  
As this function is locally analytic  the Fourier coefficients decay rapidly at large $q$ and the rate
of decay is related to the width of analytical continuation in the complex plane.
This function can be analytically continued on the complex plane in the region $\alpha < |z| < \alpha^{-1}$ with 
$f(z) = (1  + \alpha^2 - \alpha(z + z^{-1})^{-s} =  
\frac1{2}   \sum_{n= -\infty}^\infty b_s^n (\alpha) z^n$.   The Cauchy root test for convergence
implies that in the limit of large $n$ that $|b_s^n(\alpha)| \lesssim \alpha^n$ and so the Fourier 
coefficients decay rapidly.   
We approximate Laplace coefficient with the function  
\begin{equation}
b_{1/2}^{(q)} (\alpha) \sim 0.5 |\ln \delta| \exp(- \delta |q|)
\label{eqn:lapapprox}
\end{equation}
where $\delta$ is the interplanetary separation (equation \ref{eqn:delta}) and 
$\delta = \alpha^{-1} -1 \approx 1 - \alpha$. 
This Laplace coefficient diverges logarithmically for small $\delta$  (or for $\alpha$ near 1) and drops exponentially 
at large $q$.   
In  Figure \ref{fig:lap} we graphically show this approximation for the Laplace coefficient for $\delta$ in the range 0.2 to 0.01.

We can write the interaction term as a function of Poincar\'e coordinates.
\begin{equation}
W_{ij,q} =  -{m_j^3 m_i  \over  \Lambda_j^2} 
       b_{1/2}^{(q)} \left({\Lambda_i^2 m_j^2 \over \Lambda_j^2 m_i^2}\right). % \cos (q \lambda_j - q\lambda_i).
\end{equation} 
%with 
%\begin{equation}
%{\alpha_{ij}} \equiv {a_i \over a_j} = \left({\Lambda_i m_j \over \Lambda_j m_i}\right)^2.
%\end{equation}

Indirect terms can be neglected here because they only contribute a single zero-th order
Fourier component, that with $q=1$ (e.g., see appendix Table B.2 by \citealt{M+D}).

\section{Three body resonances}

We  consider the possibility that the system is not in any two body resonances where
$p n_i \sim q n_j$ with integers $p,q$, but might be in a three body resonance.
Here $n_i$ is the mean motion of the $i$-th body.
A Laplace relation exists between three orbiting bodies if the frequency
\begin{equation}
 p n_i - (p+q) n_j + q n_k \sim 0. 
\label{eqn:Laplacef}
\end{equation}
Integrating the previous equation (and assuming that the precession rates
are slow) we find that the angle
\begin{equation}
\phi \equiv p \lambda_i - (p+q) \lambda_j + q \lambda_k \sim {\rm constant}.
\label{eqn:Laplacea}
\end{equation}
This angle
librates about a particular value (often 0 or $\pi$) when in a three-body resonance.
The period, $T$, corresponds to the time between successive 
repetitions of the initial configuration
\begin{equation}
{T \over 2\pi} \sim
{p \over n_j - n_k} \sim {q \over n_i - n_j} 
\sim { p+q \over n_i - n_k} 
\label{eqn:laplaceT}
\end{equation}
We try to maintain the ordering $a_i < a_j < a_k$ and $n_i > n_j > n_k$.

We explore how 3-body interaction terms involving angles such as given 
in equation (\ref{eqn:Laplacea}) can be constructed from 
individual 2-body interaction terms.  A similar procedure has been used before
to estimate 3-body resonance strengths \citep{aksnes88,nesvorny98,murray98,nesvorny99,guzzo05}
(also see \citealt{chirikov79}).   Our procedure is to carry out a first order canonical transformation that is 
designed to remove the two 2-body interaction terms.  The procedure is described, for example,
in section 4.1 by \citet{ferraz07}.

\subsection{Canonical transformation removing first order (in mass) two-body interaction terms}

We begin with the Hamiltonian for three bodies, lacking indirect terms and with two Fourier components from two separate two-body interaction terms.  We first consider zero-th order components (in eccentricity) so we need only consider the Poincar\'e coordinates $\vec \Lambda, \vec \lambda$ 
where the vectors refer to the the coordinates and momenta for three planets.  
We chose two components with arguments (angles) whose difference is equal to 
 a three-body Laplace angle  (equation \ref{eqn:Laplacea})
\begin{displaymath}
p(\lambda_i - \lambda_j) - q(\lambda_j-\lambda_k) =  p \lambda_i - (p+q)\lambda_j + q\lambda_k.
\end{displaymath}
Our Hamiltonian with these two components
\begin{eqnarray}
H(\vec \Lambda, \vec \lambda) &=& - \sum_{l=i,j,k} {m_l^3 \over 2\Lambda_l^2}  + \qquad \qquad \\
&&  W_{ij,p}  \cos (p \lambda_i - p\lambda_j) + W_{jk,q} \cos (q \lambda_j - q\lambda_k) \nonumber 
\end{eqnarray}
with the functions
\begin{eqnarray}
W_{ij,p}(\Lambda_i,\Lambda_j) & = & - {m_i m_j^3  \over  \Lambda_j^2} b_{1/2}^{(p)} (\alpha_{ij}) \nonumber \\ 
W_{jk,q}(\Lambda_j,\Lambda_k)& = &- {m_j m_k^3  \over  \Lambda_k^2} b_{1/2}^{(q)} (\alpha_{jk}) 
\end{eqnarray}
as discussed in the previous section.
We have used the  notation
\begin{eqnarray}
\alpha_{ij}  &\equiv & {m_j^2 \Lambda_i^2 \over m_i^2 \Lambda_j^2}  = {a_i \over a_j} \nonumber \\
\alpha_{jk} &\equiv & {m_k^2 \Lambda_j^2 \over m_j^2 \Lambda_k^2} = {a_j \over a_k}. \nonumber
\end{eqnarray}

It is useful to compute some derivatives
\begin{eqnarray}
{dW_{ij,p} \over d\Lambda_i} &=& -{2 m_i m_j^3 \over \Lambda_i \Lambda_j^2 }  \alpha_{ij} D_\alpha  b_{1/2}^p(\alpha_{ij}) \nonumber \\
{dW_{ij,p} \over d\Lambda_j} &=& {2 m_i m_j^3 \over \Lambda_j^3} \left[ 1 + \alpha_{ij} D_\alpha \right] b_{1/2}^p(\alpha_{ij}) \nonumber \\
{dW_{jk,q} \over d\Lambda_j} &=& -{2 m_j m_k^3 \over \Lambda_j \Lambda_k^2 }  \alpha_{jk} D_\alpha  b_{1/2}^q(\alpha_{jk}) \nonumber \\
{dW_{jk,q} \over d\Lambda_k} &=& {2 m_j m_k^3 \over \Lambda_k^3} \left[ 1 + \alpha_{jk} D_\alpha \right] b_{1/2}^q(\alpha_{jk}) 
\end{eqnarray}
with $D_\alpha \equiv {d \over d \alpha}$.
We can use the shorthand
\begin{eqnarray}
n_{ij} \equiv n_i - n_j \qquad \lambda_{ij} \equiv \lambda_i - \lambda_j
\end{eqnarray}
and similarly with indices $jk$.

We use a generating function with new momenta $\vec \Lambda'$ and old coordinates $\vec \lambda$
\begin{eqnarray}
F_2(\vec \Lambda',\vec \lambda) &=& \sum_{l=i,j,k} \Lambda_l' \lambda_l 
- {W_{ij,p}' \over p n_{ij}' } \sin (p \lambda_{ij})\nonumber \\
&& - {W_{jk,q}' \over q n_{jk}' } \sin (q \lambda_{jk} )
\end{eqnarray}
to generate a canonical transformation.  Here $W_{ij,p}'$ is a function of $\Lambda_i' $ and $\Lambda_j'$
and similarly $n_{ij}'$.  
This canonical transformation is designed to remove the two perturbation terms in the Hamiltonian 
to first order in the planet masses. 
Derivatives of the generating function give us new coordinates in terms of the old ones
\begin{eqnarray}
\Lambda_i &=& \Lambda_i' - {W_{ij,p}'  \over  n_{ij}' }  \cos (p\lambda_{ij})    \\
\Lambda_j &=& \Lambda_j' + {W_{ij,p}' \over  n_{ij}' }   \cos (p\lambda_{ij} )
                                            - {W_{jk,q}' \over  n_{jk}' } \cos (q\lambda_{jk} ) \nonumber \\
\Lambda_k &=& \Lambda_k' +{ W_{jk,q}'  \over  n_{jk}'} \cos (q\lambda_{jk})  \nonumber \\
\lambda_i'  &=& \lambda_i + \left[{\partial n_i' \over \partial \Lambda_i'} {W_{ij,p}'  \over p n_{ij}'^2}  
                        - {\partial W_{ij,p}' \over \partial \Lambda_i'} { 1 \over p n_{ij}'} \right] \sin (p\lambda_{ij}) \nonumber \\
\lambda_j' &=&  \lambda_j - \left[{\partial n_i' \over \partial \Lambda_j'} {W_{ij,p}'  \over p n_{ij}'^2}  
                        + {\partial W_{ij,p}' \over \partial \Lambda_j'}  { 1 \over p n_{ij}'} \right] \sin (p\lambda_{ij}) \nonumber \\
                 &&       + \left[{\partial n_j' \over \partial \Lambda_j'} {W_{jk,q}'  \over q n_{jk}'^2}  
                        - {\partial W_{jk,q}' \over \partial \Lambda_j'}  { 1 \over q n_{jk}'} \right] \sin (q\lambda_{jk}) \nonumber \\
\lambda_k' &=& \lambda_k -\left[ {\partial n_k' \over \partial \Lambda_k'} {W_{jk,q} ' \over q n_{jk}'^2}  
                        + {\partial W_{jk,q}' \over \partial \Lambda_k'}  { 1 \over q n_{jk}'} \right] \sin (q\lambda_{jk}). \nonumber
\end{eqnarray}
It is useful to relate
\begin{equation}
{\partial n_i \over \partial \Lambda_i} = - {3 m_i^3 \over \Lambda_i^4}
\end{equation}
and similarly for the other bodies.

We replace our old coordinates and momenta with new ones in the Hamiltonian finding that in the new coordinates the first order two body terms have been removed by the transformation.  We expand to second order in the masses and find that the Hamiltonian has gained second order terms.   During this procedure we drop terms
that depend on  $\cos^2(p \lambda_{ij})$, or $\sin^2 (p \lambda_{ij})$ and $\cos^2(q \lambda_{jk})$,  or $\sin^2(q \lambda_{jk})$ while keeping those containing the products $\cos (p\lambda_{ij}) \cos (q\lambda_{ik})$ and 
$\sin (p\lambda_{ij}) \sin(q \lambda_{jk})$.  We rewrite these products in terms of the sum and difference of the angles and discard the term that contains the sum of the angles so as to retain only the term that depends on the Laplace angle $q \lambda_i - (p+q) \lambda_j +q \lambda_k$.
When the Laplace angle is slowly varying (and the system is near a three-body resonance)
the other terms are rapidly varying and so can be neglected.
\begin{equation}
H(\vec \Lambda', \vec \lambda') %&=& H_0 + \epsilon_{pq} \cos (p \lambda_i - (p+q)\lambda_j + q \lambda_k)  ~~~~ \\
= -\sum_{l=i,j,k}  {m_l^3 \over 2\Lambda_l'^2} + \epsilon_{pq} \cos (p \lambda_i' - (p+q)\lambda_j' + q \lambda_k') 
\label{eqn:Ham}
\end{equation}
with
\begin{eqnarray}
%H(\vec \Lambda',\vec \lambda') &\approx & \sum_{l=i,j,k}  -{m_l^3 \over 2\Lambda_l'^2}  +
\epsilon_{pq} &\approx&  {m_i m_j m_k^3 \over \Lambda_k'^2} {  \bigg [  } \nonumber \\
&&  {3 n_j'^2\over 2 } 
	\left( {1 \over 2n_{ij}' n_{jk}'} + {p \over q n_{ij}'^2} + {q \over p n_{jk}'^2} \right) b_{1/2}^p(\alpha_{ij}') b_{1/2}^q (\alpha_{jk}')  \nonumber \\
&& + \left( {n_j' \over n_{jk}'} + {q n_j' \over p n_{ij}'} \right) b_{1/2}^q(\alpha_{jk}')\left( 1 + \alpha_{ij}' D_\alpha \right) b_{1/2}^p (\alpha_{ij}')   \nonumber \\
&& +  \left( {n_j' \over n_{ij}'} + {p n_j'\over q n_{jk}'} \right)b_{1/2}^p (\alpha_{ij}') \alpha_{jk}' D_\alpha b_{1/2}^q( \alpha_{jk}')  {\bigg ]} % \nonumber \\
%&& \times \cos (p \lambda_i' - (p+q)\lambda_j' + q \lambda_k'). 
\label{eqn:ick}
\end{eqnarray}
Our procedure using a canonical transformation should give a resonance term equivalent to the zero-th order term derived by \citet{aksnes88}  using Lagrange's equations, though it is not easy to check because of the differences in notation.
The resonance term is zero-th order in planet eccentricity but second order in the planet masses.
Hereafter we drop the primes in the coordinates.

The full Hamiltonian contains additional two-body terms however
the angles involved are expected to vary quickly compared to $\phi$.  
Neglecting fast angles is equivalent to averaging over them.
Equivalently as long as there are no combinations that yield slow angles, the other interaction terms may be removed using near identity canonical transformations similar to that used above.

\subsection{Width, libration frequency and conserved quantities for the zero-th order three-body resonance}

The Hamiltonian (equation \ref{eqn:Ham}) that contains a three-body term can be used to estimate the width and timescales in a Laplace resonance.   
We can perform a canonical transformation to reduce the dimension of the problem.
Consider the generating function
\begin{equation}
F_2(\vec \lambda, \vec J) = (p \lambda_i - (p+q) \lambda_j + q \lambda_k)J + \lambda_j J_j + \lambda_k J_k
\end{equation}
leading to new angles $( \phi, \lambda_j', \lambda_k')$
\begin{eqnarray}
\phi &=& (p \lambda_i - (p+q) \lambda_j + q \lambda_k)  \nonumber \\
 \lambda_j' &=& \lambda_j  \nonumber \\ \lambda_k' &=& \lambda_k 
 \end{eqnarray}
 and new momenta $(J, J_j, J_k)$ such that
 \begin{eqnarray}
pJ &=&  \Lambda_i  \nonumber \\ 
-(p+q) J + J_j &=& \Lambda_j  \nonumber   \\   
q J + J_k &=& \Lambda_k.  
\label{eqn:pJ}
\end{eqnarray}

After the transformation, the new Hamiltonian (using equation \ref{eqn:Ham}) is
\begin{eqnarray}
K(J,J_j, J_k; \phi, \lambda_j, \lambda_k) &=& - {m_i^3 \over 2 p^2 J^2} - {m_j^3 \over 2( J_j - (p+q)J)^2} \nonumber \\
&& - {m_k^3 \over 2 (J_k + q J)^2} + \epsilon_{pq} \cos \phi.
\label{eqn:K2}
\end{eqnarray}
The new Hamiltonian only depends
on the angle $\phi$ and  does not depend on the two longitudes $\lambda_j, \lambda_k$
(that are unchanged by the canonical transformation)
so the two conjugate momenta $J_j, J_k$ are conserved quantities.  
Our conserved quantities can be written as 
\begin{eqnarray}
p J_j &=& p \Lambda_j + (p+q) \Lambda_i  \nonumber \\
pJ_k &=& p \Lambda_k - q \Lambda_i.  \label{eqn:conserved}
\end{eqnarray}
Differentiating these conserved quantities with respect to the semi-major axes we find that small changes
\begin{eqnarray}
{da_k \over d a_i} &=& {m_i \over m_k}{q \over p} \left({a_k \over a_i} \right)^{1/2} \nonumber \\
{da_j \over d a_i} &=& -{m_i \over m_j}{ (p+q) \over p} \left({a_j \over a_i} \right)^{1/2} .
\label{eqn:da}
\end{eqnarray}
These derivatives imply that a small change in semi-major axis by one body will be mirrored by
changes in semi-major axis of the two other bodies.
The signs imply that the outer two bodies move in the same direction but the middle one moves in the opposite
 direction.  The middle body is expected to move more than the outer two as we expect $p+q$ is greater than $p$ and $q$.  This behavior can be seen in particle integrations as we will discuss below. 

We can expand the momentum $J$ about an initial value.   Consider initial values for $\Lambda_{i0}, \Lambda_{j,0}, \Lambda_{k0}$ corresponding to initial value  $J_0$, conserved quantities $J_{j0}, J_{k0}$, initial semi-major axes
and mean motions
\begin{eqnarray}
{\bf a}_0 = (a_{i0}, a_{j0}, a_{k0}) \qquad {\bf n_0}  = (n_{i0},n_{j0},n_{k0}). \label{eqn:ndef}
\end{eqnarray} 
 We define
\begin{equation} J \equiv  J_0 + I, \end{equation}
and expand the Hamiltonian (equation \ref{eqn:K2}) about $J_0$. To second order in $I$
\begin{equation}
K(\phi, I) = A {I^2\over 2} + B I + \epsilon_{pq} \cos \phi + {\rm constant}
\label{eqn:Kam}
\end{equation}
where the constant contains terms that depend on our conserved quantities ($J_j, J_k$) and $J_0$.
The coefficients
\begin{eqnarray}
B &=& p n_{i0} - (p+q) n_{j0} + q n_{k0}  \nonumber \\
A &=& -{3 }\left({p^2 \over m_i a_{i0}^2} + {(p+q)^2\over m_j a_{j0}^2} + {q^2 \over m_k a_{k0}^2}\right). 
\label{eqn:AB}
\end{eqnarray}
The coefficient $\epsilon_{pq}$ is also evaluated at ${\bf a}_0$.
%and 
%\begin{equation} c = {3 (p+q) J_j \over a_j^2 m_j} - {3 q J_k \over a_k^2 m_k} \end{equation} should be small.
We can think of the coefficient $B$ as a product 
\begin{equation}
B = {\bf z}  \cdot {\bf  n_0} \label{eqn:B}
\end{equation}
setting distance to resonance, where the vector of integers 
\begin{equation} {\bf z} \equiv (p, -(p+q), q). \label{eqn:zdef}
\end{equation}
On resonance $B\sim0$.  The coefficient $A$  depends approximately
on the magnitude of the vector ${\bf z}$ with $|A| \sim |{\bf z}|^2/(m a_{i0}^2)$.

With a shift in the momentum, $I_s = I + B/A$, the Hamiltonian (equation \ref{eqn:Kam}) can be written
so as to remove the term that is proportional to $I_s$,
\begin{equation}
K(\phi, I_s) = A {I_s^2\over 2} + \epsilon_{pq} \cos \phi + {\rm constant}.
\end{equation}
We estimate the width of the resonance in momentum is
\begin{equation}
\Delta I \sim 2 \sqrt{2\epsilon_{pq}\over A}
\label{eqn:Iai}
\end{equation}
corresponding to a resonant width in Poincar\'e momentum $\Lambda_i$ of
\begin{equation}
\Delta \Lambda_i \sim 2 p \sqrt{2\epsilon_{pq}\over A}
\label{eqn:Lai}
\end{equation}
(using equation \ref{eqn:pJ})
and a width in terms of semi-major axis of the innermost body
\begin{equation}
\Delta a_i \sim {4 p \over m_i} \sqrt{2\epsilon_{pq} a_i \over A}.
\label{eqn:Dai}
\end{equation}
A jump across resonance would give a change $\Delta a_i$ for the inner body.
Using equations (\ref{eqn:da}) changes in semi-major axis for the other two bodies can
be estimated from that of the inner one.

The libration frequency in the resonance is 
\begin{equation} 
\omega_{pq} \sim \sqrt{\epsilon_{pq} A}. 
\label{eqn:omega}
\end{equation}
For a system initially with $I=0$ to be in resonance we require that the shift, $B/A$, (relating
$I$ and $I_s$) is smaller
than half the resonance width ($\Delta I/2$ in equation \ref{eqn:Iai}).  
Using this condition and equations (\ref{eqn:Iai}) and
 (\ref{eqn:omega}) we require
$|B| \lesssim \sqrt{2} \omega_{pq}$ to be near or in resonance  or using equation (\ref{eqn:B})
\begin{equation}
|{\bf z} \cdot {\bf n}_0| \lesssim \sqrt{2} ~\omega_{pq}.
\label{eqn:inres}
\end{equation}

\subsection{Estimates of three-body resonance strengths and frequencies for 
 closely and evenly spaced equal mass multiple planet systems}

We consider the strength of the various terms contributing to the Laplace resonance strength, $\epsilon_{pq}$
in equation (\ref{eqn:ick}), for a closely and evenly spaced equal mass system.
For the equally spaced system (represented by equation \ref{eqn:delta})
 $\delta_{ij} \sim \delta_{jk}$.    In this setting the only Laplace angles that can be nearly fixed have $p$ about
 the same size as $q$. We work in units of the mean motion and semi-major axis of the innermost body
 involved in the three-body resonance.
Differences in the mean motions can be approximated as 
\begin{equation} n_{ij} \sim {3 \over 2}\delta_{ij}\end{equation} where 
\begin{equation} \delta_{ij} \equiv 1 - \alpha_{ij}.\end{equation}  
As the Laplace coefficient $b_{1/2}^{(q)}(\alpha)$ can be approximated given in equation (\ref{eqn:lapapprox}),   
the derivatives of the Laplace coefficients can be approximated as 
\begin{equation} 
D_\alpha b_{1/2}^{(q)} (\alpha) \sim 0.5 (\delta^{-1} + |q \ln \delta|) \exp(-\delta |q|)
 \end{equation}
(using $D_\alpha = - {d \over d\delta}$).
%The terms containing the derivatives of the Laplace coefficients have the same sign of the other terms.   
Using these approximations and assuming equal masses,
we find that the interaction term strength in equation (\ref{eqn:ick}) is approximately
\begin{eqnarray}
\epsilon_{pq} &\sim& m^3 \left[  \delta^{-2} (\ln \delta)^2  
+  0.5 \delta^{-2}\left(2 + (p +q)\delta |\ln \delta)| \right) | \ln \delta| \right]\nonumber  \\
&& \times \exp(-\delta (p+q)).
\end{eqnarray}
where we have assumed integers $p,q$ are positive.
The terms all have the same sign and in most cases
the first term dominates so we can approximate the interaction strength as
\begin{equation}
\epsilon_{pq} \sim m^3  \delta^{-2} (\ln \delta)^2 \exp(-\delta (p+q)) .  
\label{eqn:epsilon}
\end{equation}
Note that $\epsilon_{pq}$ is positive, so we would expect libration around Laplace angle of zero or
\begin{equation}
\phi = p \lambda_i - (p+q) \lambda_j + q \lambda_k \approx 0.
\end{equation}

For $p \sim q$ (corresponding to being near resonance for small $\delta$) 
we estimate that $|A| \sim 20 p^2/m$ (using equation \ref{eqn:AB}) and from equation (\ref{eqn:omega}) 
and equation (\ref{eqn:epsilon}) the libration frequency in the resonance is approximately
\begin{equation}
\omega_{pq} \sim 4 m p \delta^{-1} |\ln \delta| \exp(- \delta p). 
\label{eqn:om}
\end{equation}

The width of the resonance (in terms of momentum $\Lambda_i$; equations \ref{eqn:Lai}, \ref{eqn:Dai}) 
gives a width in  semi-major axis similar in size or
\begin{equation}
\Delta a_i \sim  m \delta^{-1} |\ln \delta| 
\label{eqn:dai2}
\end{equation}
where we have assumed $p\delta  \lesssim 1$ and so neglected the exponential.
The above equation should give the size of jumps across resonance for the inner body.

% comments
Likely equation (\ref{eqn:dai2}) somewhat underestimates the resonance widths as we have not taken
into account all terms in equation (\ref{eqn:ick}).  We have also neglected the dependence 
of $\epsilon_{pq}$ on the momentum $J$
in our estimates of $\omega_{pq}$ (equation \ref{eqn:omega}) and $\Delta a_i$ (equation \ref{eqn:Dai}). 
The conserved quantities in resonance imply that 
$\delta_{ij}$  increases when $\delta_{jk}$ decreases and vice-versa, so $\epsilon_{pq}$ may
not be strongly dependent on $J$. 

Because the Laplace coefficients drop exponentially with separation $\delta$ (see Figure \ref{fig:lap}) 
or the distance between the planets, resonances 
between the first, second and fourth bodies or other non-consecutive combinations 
should be much weaker than those involving three consecutive bodies.  
However if the masses of the bodies
differ then three-body resonances involving non-consecutive triplets could be important.
%For the equal mass systems, the lack of dependence of crossing
%or stability timescales on the number of bodies, $N$ when $N\gtrsim 5$ \citep{chambers96}
%may be attributed to the exponential drop in resonance strengths with interplanetary separation.

\subsubsection{First order resonances}

Zero-th order (in eccentricity) resonances do not influence the Poincar\'e variable  associated with eccentricity
so they should not affect planet eccentricities.
First order three-body resonances (as previously considered by \citealt{aksnes88}) have interaction terms in the  form 
\begin{equation}
\eta \Gamma_j^{1/2} \cos(p \lambda_i - (p+q-1)\lambda_j + q \lambda_k - \varpi_l)
\label{eqn:ex}
\end{equation}
with a single longitude of pericenter $\varpi_l$.   
The longitude of pericenter can be that of any of the three planets
involved in the resonance, so that $l$ can be equivalent to $i,j$ or $k$. 
Because the interaction term contains a longitude of pericenter it can affect a planet's eccentricity.  
The strength $\eta$ can be estimated in the same way as we have estimated the zero-th order three-body 
resonance strengths but instead of carrying out a transformation 
with two zero-th order two-body interaction Fourier terms, 
one begins with a zero-th order and a first order two-body Fourier term.

\begin{figure}%[htbp]
\begin{center}
\includegraphics[width=8cm]{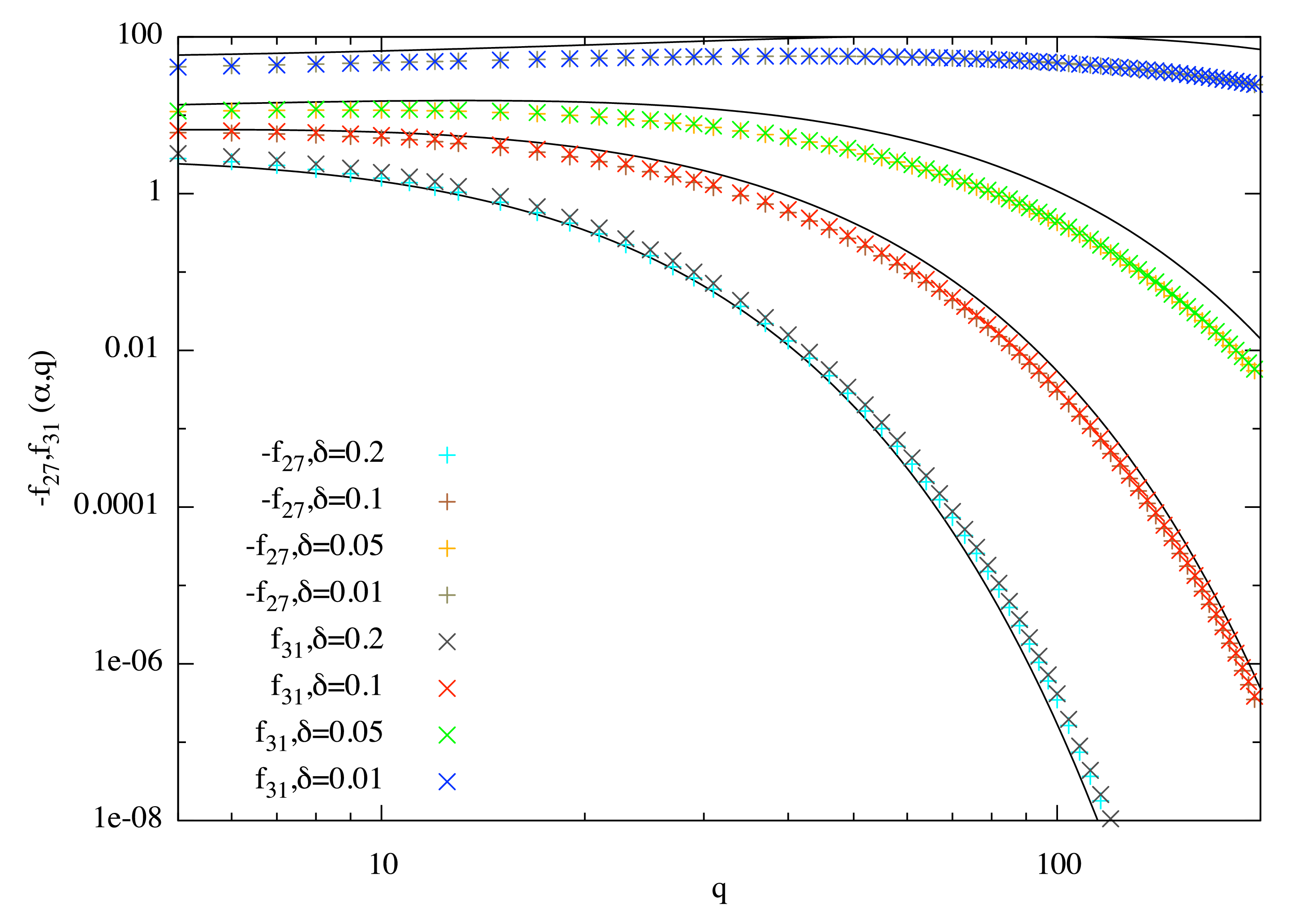}
\caption{Approximations to the coefficients $f_{27},f_{31}$. 
Plotted as points are the coefficients $-f_{27}(\alpha,q)$ and $f_{31}(\alpha,q)$ (equations \ref{eqn:f27f31}) 
as a function of the integer $q$ 
for four separations $\delta=\alpha^{-1}-1 = 0.01,0.05,0.1,$ and $0.2$.  Overplotted as thin black lines
for each $\delta$ value is the function $0.5 (\delta^{-1} + |q \ln\delta|) \exp(- \delta|q|)$.  These functions
are approximately equivalent to $D_\alpha b_{1/2}^{(q)}(\alpha)$.
 }
\label{fig:lap1}
\end{center}
\end{figure}

When expanded to first order in planet eccentricity and inclination the two-body interaction terms 
(equation \ref{eqn:Wij})
gain Fourier components (that would be added to equation \ref{eqn:Wijf})
\begin{eqnarray}
\sum_{q=-\infty}^\infty 
\left[ 
  	V_{ij,q}^a  \cos (q \lambda_j + (1-q)\lambda_i - \varpi_i)+  \qquad \qquad \right. \nonumber \\
	\qquad \qquad   \left. V_{ij,q}^b \cos (q \lambda_j + (1-q)\lambda_i - \varpi_j)
\right]
\label{eqn:Vsum}
\end{eqnarray}
where 
\begin{eqnarray}
V_{ij,q}^a &= - {m_i m_j \over a_j} e_i f_{27}(\alpha_{ij},q) \approx
-{m_i m_j^3 \over \Lambda_j^2} \left(2\Gamma_i \over  \Lambda_i \right)^{1\over 2} f_{27} (\alpha_{ij},q)    \nonumber \\
 V_{ij,q}^b &=  -{m_i m_j \over a_j} e_j f_{31}(\alpha_{ij},q) \approx
 -{m_i m_j^3 \over \Lambda_j^2}\left(2\Gamma_j \over  \Lambda_j \right)^{1\over 2} f_{31} (\alpha_{ij},q)
 \nonumber \\
\end{eqnarray}
and coefficients
\begin{eqnarray} 
f_{27}(\alpha,q) &\equiv& {1 \over 2}\left[ -2 q - \alpha D_\alpha \right] b_{1/2}^{(q)}(\alpha)\nonumber \\  
f_{31}(\alpha,q) &\equiv& {1 \over 2} \left[ -1 + 2 q + \alpha D_\alpha  \right] b_{1/2}^{(q-1)}(\alpha)
\label{eqn:f27f31}
\end{eqnarray}
(equation 6.107 \citealt{M+D}; also see Tables B.4 and B.7).
An approximation to these coefficients is $-f_{27} \sim f_{31} \sim 0.5 (\delta^{-1} + |q \ln\delta| )\exp(-\delta |q|)$
shown in Figure \ref{fig:lap1}.    Also $-f_{27} \sim D_\alpha b_{1/2}^{(q)}(\alpha)$.
% While the sum is for all j even negative ones, the ordering chosen here is a_j > a_i so that n_j < n_i
% and we only have positive j terms being important in the above two expressions when we are only
% looking at two-body things, 
%but remember than the opposite sign ones are of interest for the three body resonances, 

First order and zero-th order terms when combined form a term with a three body argument similar
to that shown in equation (\ref{eqn:ex}).
We list below on the left the two Fourier components that when combined give the argument on the right;
\begin{eqnarray}
V_{ij,-p}^a(\alpha_{ij}) W_{jk,q}(\alpha_{jk})&:& (p+1)\lambda_i - (p+q)\lambda_j + q \lambda_k - \varpi_i   \nonumber \\
V_{ij,-p}^b(\alpha_{ij}) W_{jk,q}(\alpha_{jk}) &:&  (p+1)\lambda_i - (p+q)\lambda_j + q \lambda_k - \varpi_j   \nonumber \\
V_{jk,q}^a(\alpha_{jk}) W_{ij,p}(\alpha_{ij}) &:&  p\lambda_i - (p+q-1)\lambda_j + q \lambda_k - \varpi_j     \nonumber \\
V_{jk,q}^b(\alpha_{jk}) W_{ij,p}(\alpha_{ij}) &:& p\lambda_i - (p+q-1)\lambda_j + q \lambda_k - \varpi_k . \nonumber \\
&
\end{eqnarray}
The longitude of pericenter in the argument determines the eccentricity related Poincar\'e coordinate.  
For example, if the argument contains $\varpi_i$
then the three-body term contains a factor of $\Gamma_i^{1/2}$ and similarly for bodies $j,k$.

Whereas the zero-th order three-body resonances
involve two $W$ coefficients, the first order three-body resonances involve a single $W$ and a single $V$ coefficient.   The $V_{ij}$ coefficients are approximately derivatives of the $W_{ij}$ coefficients. 
The strength, $\eta$, of the first order three-body term we expect is larger than $\epsilon_{pq}$ by a factor
of $\delta^{-1}$  because it would involve an extra derivative of the Laplace coefficient. 
However the dependence on the Poincar\'e coordinate $\Gamma$
associated with the eccentricity can reduce the strength of the resonance.
The first few conjunctions of an initially zero eccentricity system lead to planet eccentricities of order a few times 
$m \delta^{-2}$ (e.g., equation 10.57 \citealt{M+D}).
This suggests that the ratio of the first to zero-th order resonance strengths is approximately 
$\eta \Gamma_j/\epsilon_{pq} \sim m \delta^{-3}$.  It may be convenient to define $\delta_h = \delta/r_h$
where $r_h \propto m^{1/3}$ is the Hill radius of the planet.   The ratio of resonance strengths is
then approximately $3 \delta_h^{-3}$,  or the interplanetary separation in units of the planet's Hill radius.
For eccentricities above a critical value (e.g., set dimensionally; \citealt{quillen06})  
the resonant width and libration frequency depends  on the square root of $\eta \Gamma_j$. 
%(though at low eccentricity they depend on $\eta$ to the 2/3 power).
As $\delta_h$ is in the range $\sim 2$--10 for the systems studied numerically (e.g., \citealt{chambers96,smith09})
we expect that first order resonances are initially a few times weaker than the zero-th order ones.
However, a system evolving in multiple three-body zero-th order resonances can cross 
first order resonances causing variations in planet eccentricities. Because of the number
of possible angular combinations there are more first order resonances than zero-th order ones.

\section{Three-body resonances as seen in a numerical integration}

\begin{figure}%[htbp]
\begin{center}
\includegraphics[width=8cm]{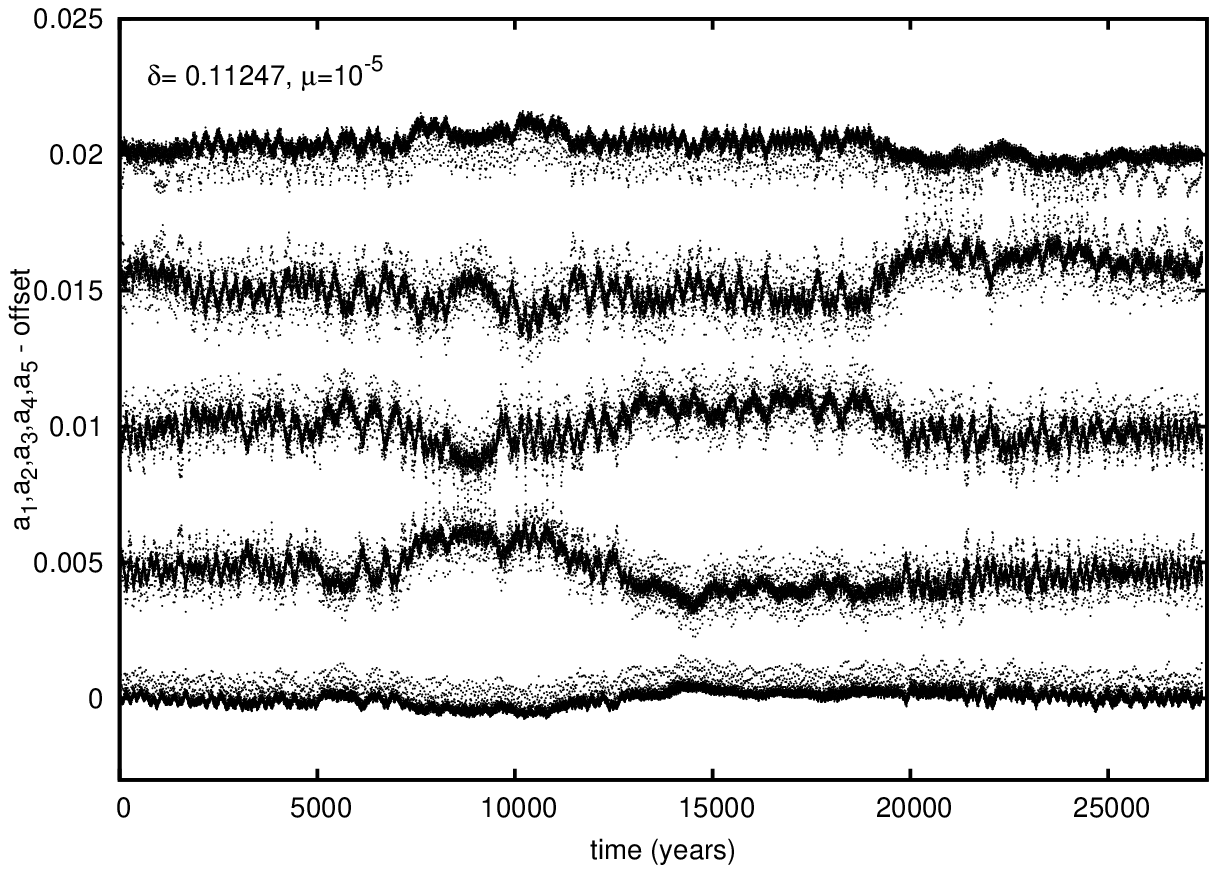}
\includegraphics[width=8cm]{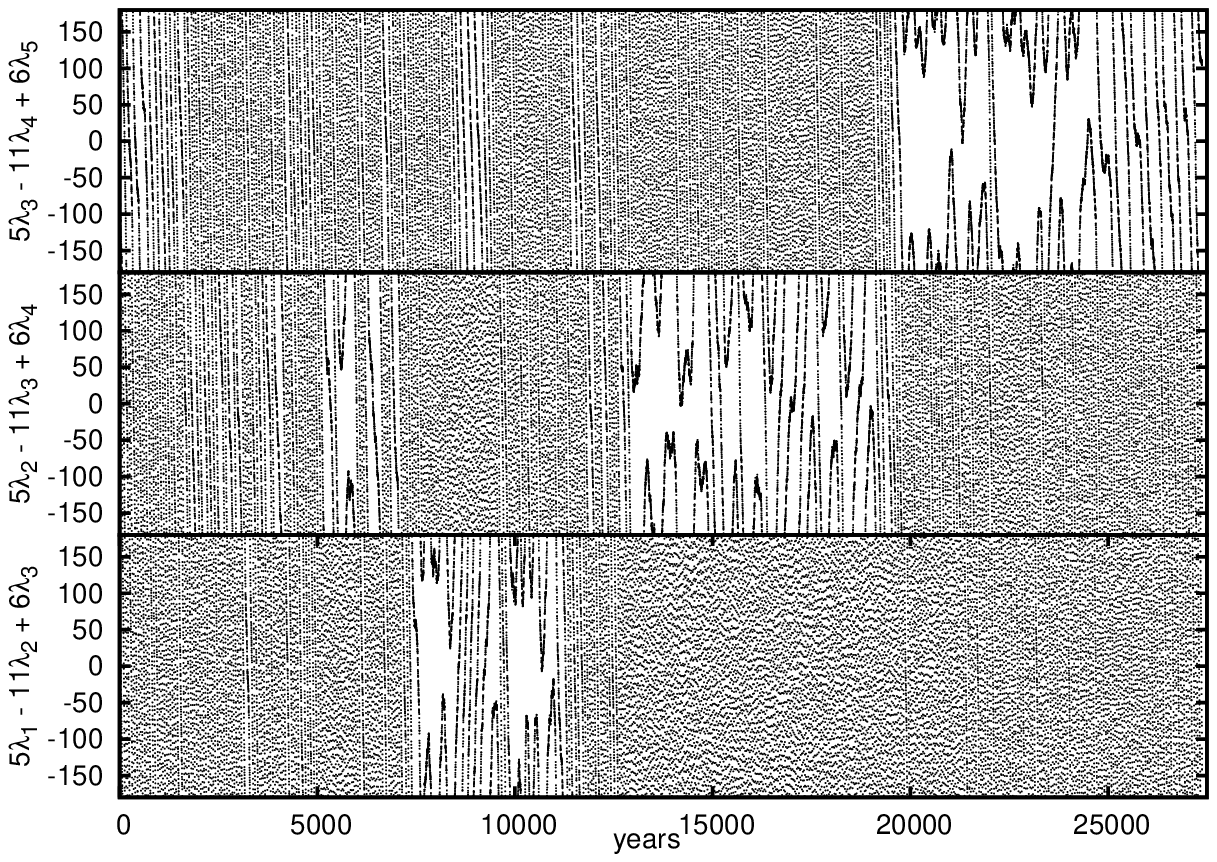}
\includegraphics[width=8cm]{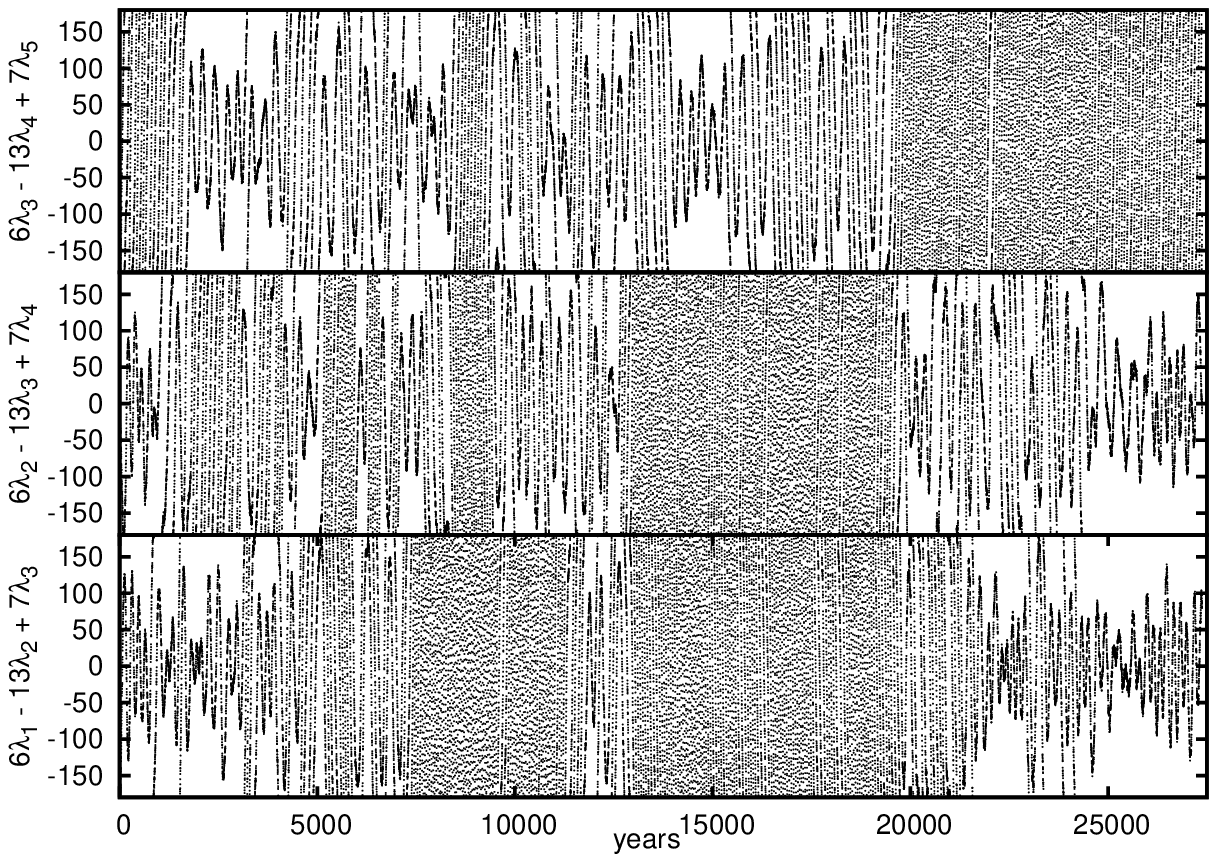}
\caption{An example of a numerical integration of 5 equal mass tightly packed bodies experiencing three body resonance crossings.  
For this simulation the mass ratio $m=10^{-5}$ and initial interplanetary spacing $\delta = 0.11247$.
 a) The semi-major axes as a function of time in rotation periods of the innermost body are shown for all 5 bodies.  Each set of points has been shifted by an arbitrary amount but has not been rescaled.  
 b) We show Laplace angles $\phi = p \lambda_i - (p+q) \lambda_j + q \lambda_k$ in degrees as a function of time for $p=5, p+q=11,q=6$ (or ${\bf z} = (5,-11,6)$ for the inner three consecutive bodies 
 (bottom panel; with planet indices $i=1,j=2,k=3$), the middle three bodies 
 (middle panel; $i=2,j=3,k=4$) and the outer three consecutive bodies (top panel).  
 c) Similar to b) except for resonances with $p=6,q=7$ (or ${\bf z} = (6,-13,7)$. 
When one of Laplace angle ceases to circulate and librates about 0 or $\pi$, variations in the semi-major axis
in three of the bodies are related by two conserved quantities (equation \ref{eqn:conserved}, \ref{eqn:da}).  
The middle planet moves opposite to the outer two planets and the middle planet experiences larger variations in semi-major axis than the other two.  
}
\label{fig:threejump}
\end{center}
\end{figure}

In Figure \ref{fig:threejump} we show a numerical integration of 5 equal mass bodies initially in
a coplanar circular orbits about a central star.  
The ratio of the planet masses to that of the central star is $m=10^{-5}$. 
The initial separation between the bodies is given by $\delta = 0.11247$ using 
equation (\ref{eqn:delta}).   The numerical integration was done using the hybrid algorithm of the code
Mercury version 6.2 \citep{chambers99}.  Time is given in units of the initial rotation period of the inner body. 
Distances are given in units of the inner body's initial semi-major axis.
This numerical integration was chosen to 
illustrate phenomena associated with three-body resonances and we will use it to check
predicted sizescales for them.  
For these parameters $\delta m^{-1/3} = 5.22$ and $\delta m^{-1/4} = 2.0$.
In terms of the mutual Hill radius (as defined by equation 1 by \citealt{smith09}) 
$\delta = 5.6 r_{mH}$, placing it in the middle of the regime studied by \citet{smith09} though they
primarily considered planets lower in mass by a factor of 3.  This separation places
our integration at larger separations than the mean explored by \citet{chambers96} (on the right hand side of the top panel of their Figure 3)  with a crossing timescale of about $10^5$ orbital periods.

In Figure \ref{fig:threejump}a we show the semi-major axes of the 5 bodies.
Variations in semi-major axis often involve similar motions for three consecutive bodies, with the inner and outer ones (of this triplet) moving together and the central one of the triplet moving in the opposite direction.   
This is expected as in a three-body resonance there are two conserved quantities (equations \ref{eqn:conserved})
that relate variations in semi-major axes between the three bodies (equations \ref{eqn:da}).
Figure \ref{fig:threejump}b shows the Laplace angles $\phi = p \lambda_i - (p+q) \lambda_j + q \lambda_k$
for $p=5,p+q=11,q=6$.  We can write $\phi = {\bf z} \cdot \vec \lambda$ with  ${\bf z} = (5,-11,6)$.  The angle is plotted for the inner three consecutive bodies (bottom panel), the middle three consecutive bodies (middle panel) and the outer three consecutive bodies (top panel), respectively.
Figure \ref{fig:threejump}c is similar but for the resonances with $p=6,q=7$.
We see from Figures \ref{fig:threejump}b,c that there are times when specific Laplace angles vary
slowly or vibrate about 0 or $\pi$.
The system is affected by more than one Laplace resonance.
At some times the inner three bodies move together, and at other times the middle or outer three move together.
For example,  the variations in the outer three planets at $t=19000$ periods are likely due to the ${\bf z} = (6,-13,7)$ 
resonance involving the outer three bodies.   Variations in the inner three planets at $t=9000$
are likely due to the ${\bf z} = (5,-11,6)$ resonance involving the inner three bodies.  When the Laplace angle
ceases to circulate and librates about 0 or $\pi$ we see that variations in the three planets involved
in the Laplace resonance are related, with the outer two increasing or decreasing in semi-major and the middle one moving in the opposite direction.  The middle body experiences larger variations in semi-major axis
as would be expected from the conserved quantities (see equations \ref{eqn:conserved} and \ref{eqn:da}).
We have checked that the conserved quantities in these resonances do not make large variations when
the resonant angle is not rapidly circulating.  However, conserved quantities associated with one resonance
can vary while a different resonance is affecting the system.

For $m=10^{-5}$ and  $\delta = 0.11247$ a jump across resonance should give a change in
semi-major axis for the inner body (using equation \ref{eqn:dai2}) of approximately $2 \times 10^{-4}$.
Jumps seen in the simulation (Figure \ref{fig:threejump}a) are a factor of a few larger than this.
We can consider this moderately reasonable agreement as we have only made
rough estimates of the resonance properties.
The libration frequency in the resonance (computing $\omega_{pq}$ using equation \ref{eqn:om})
is approximately $\omega_{pq} \sim 10^{-2}$ corresponding to a period of 
$T={2\pi \over \omega_{pq}}\sim 600$ years. This period is short enough that the slow variations in
angle in Figures \ref{fig:threejump}a,b can be attributed to the three-body resonances.

The simulation shown in Figure \ref{fig:threejump} is affected simultaneously by more than
one zero-th order three-body resonance (here ${\bf z} = (6,-13,7)$ and $(5,-11,6)$ resonances and for three possible  consecutive triplets of planets) suggesting that they are dense and wide enough 
that the three-body resonances overlap. 
If so then random variations in semi-major axis can be attributed to
chaotic behavior associated with multiple resonances.   

\begin{figure}
\begin{center}
\includegraphics[width=8cm]{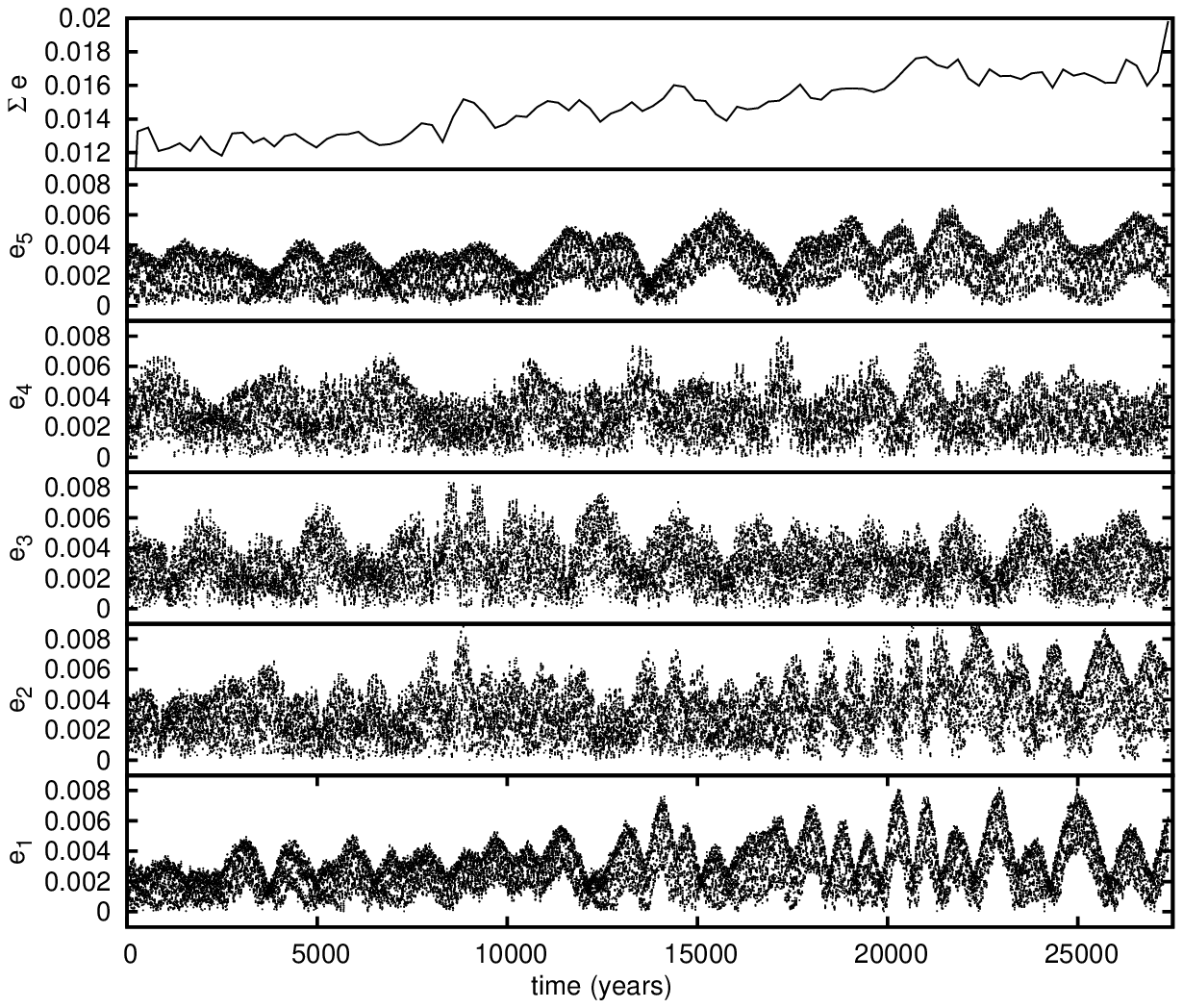}
\includegraphics[width=8cm]{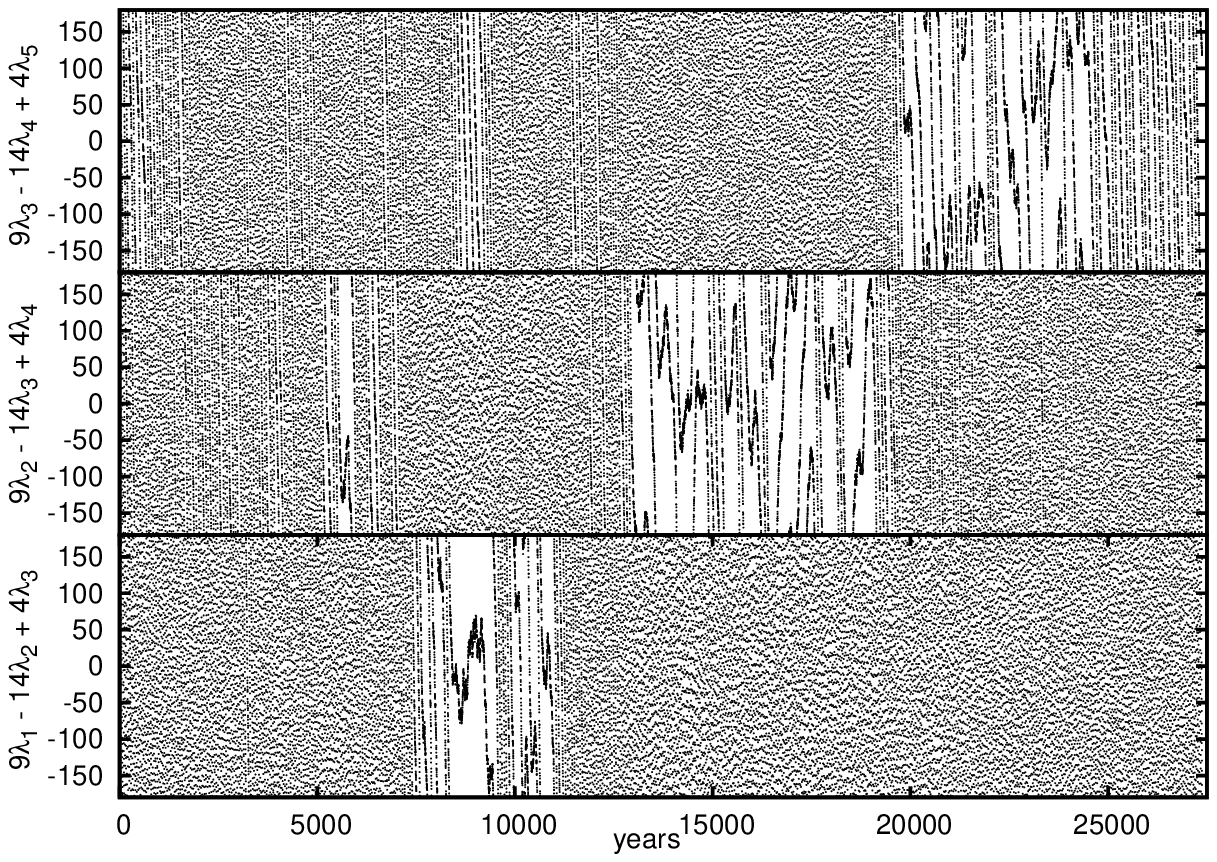}
\includegraphics[width=8cm]{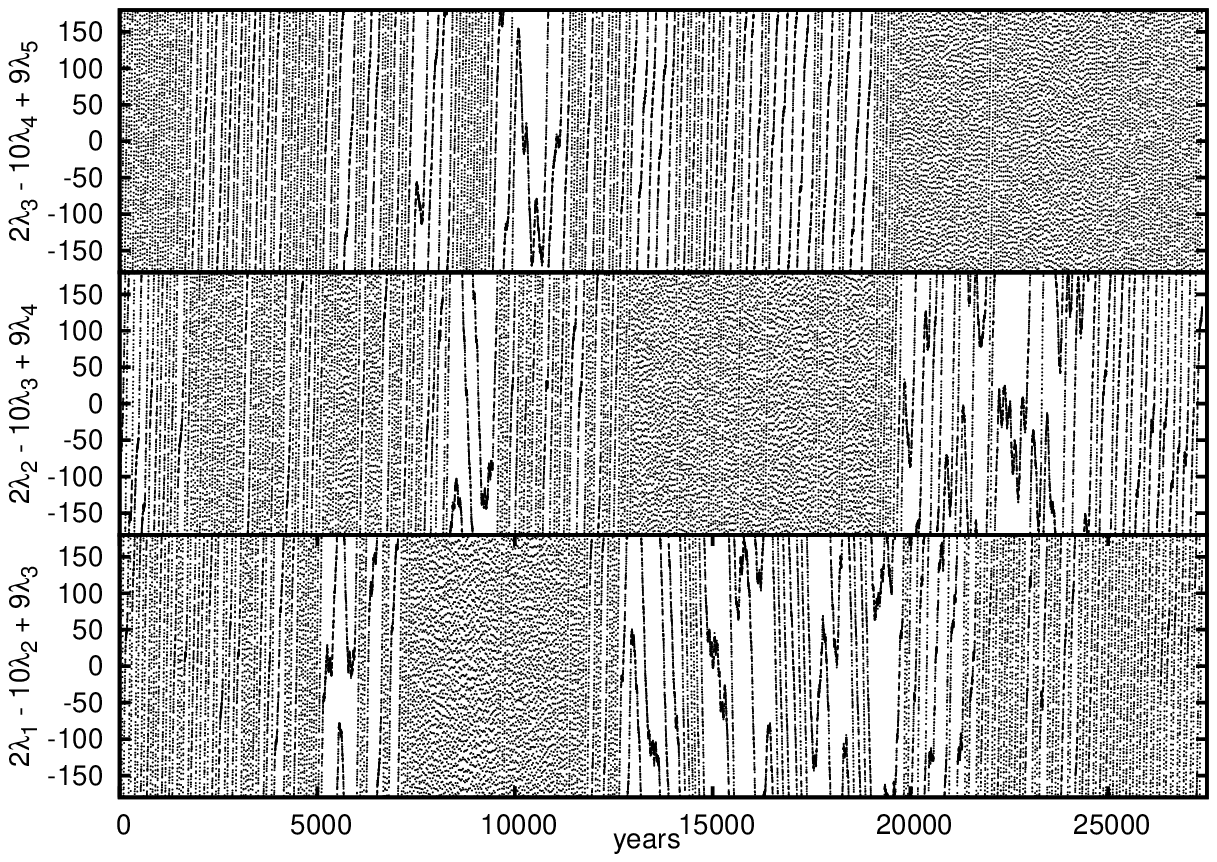}
\caption{Eccentricity evolution of the numerical integration shown in Figure \ref{fig:threejump}.  
a) The top subpanel shows the smoothed sum of the eccentricities as a function of time.  The remaining 5 subpanels show the eccentricities of the 5 planets as a function of time.  
b) We show the angles $\phi = {\bf z} \cdot \vec \lambda$ in degrees as a function of time for ${\bf z} = (9,-14,4)$ 
for the inner three consecutive bodies  (bottom panel), the middle three bodies  (middle panel), 
and the outer three consecutive bodies (top panel).  
c) Similar to b) except for ${\bf z} = (2,-10,9)$.  
When the system passes through a first order three-body resonance variations in planet eccentricities are seen.
}
\label{fig:e}
\end{center}
\end{figure}

For the same numerical integration we also show the eccentricity evolution in Figure \ref{fig:e}a.
This figure also plots angles $\phi = {\bf z} \cdot \vec \lambda$ for ${\bf z} = (9,-14,4)$ and
${\bf z} = (2,-10,9)$.   These angles are not in the form ${\bf z} =  (p, -(p+q), q)$ as the sum of
the indices in the vector $\bf z$ is not zero.    
First order (in eccentricity) thee-body resonances involve a single planet's longitude
of pericenter, for example, the angle could be one of the following 
\begin{eqnarray}
\phi &=& (p+1) \lambda_i - (p+q) \lambda_j + q \lambda_k - \varpi_i \nonumber \\
\phi &=& p \lambda_i - (p+q-1) \lambda_j + q \lambda_k - \varpi_j \nonumber \\
\phi &=& p \lambda_i - (p+q) \lambda_j + (q+1) \lambda_k - \varpi_k.  
\end{eqnarray}
These angles arise from combining a zero-th order two-body term with a first order
two-body term.  As the precession rates, $\dot \varpi$, are slow compared to the mean motions,
we have plotted angles omitting
a longitude of pericenter.   We have examined similar plots containing all of the above 
possible angular combinations
 and found that they are similar to but noisier
than Figures \ref{fig:e}b,c.  Since the planet eccentricities are low, small variations in the orbits can give
large changes in the computed longitudes of pericenter.    

From Figure \ref{fig:e}b,c we see that our numerical integrations also exhibit slow angles associated with first order three-body resonances.    When first order resonances are crossed we expect small changes in planet eccentricity. 
For example at $t\approx 8000$ the inner three planets are affected by the ${\bf z} = (9,-14,4)$ resonance
leading to an increase in eccentricity in these three planets.  At $t\approx 10,000$ years the
outer three planets are affected by the ${\bf z} = (2,-10,9)$ resonance leading to an increase
in the eccentricity of the fourth planet.  Jumps in eccentricity seem similar in size to those
 of jumps in semi-major axis though we expected them to be a few times smaller 
based on the discussion in section 3.3.1.
A comparison between the angles shown in Figure \ref{fig:threejump}b
and Figure \ref{fig:e}b shows that they are very similar; likewise for Figure \ref{fig:threejump}c and
Figure \ref{fig:e}c.   Hence the zero-th order resonances 
(with angles shown in Figure \ref{fig:threejump}b,c) overlap the first order resonances (with 
angles shown in Figure \ref{fig:e}b,c).  

Once the eccentricity of a planet is increased
secular perturbations cause oscillations in the eccentricities of the nearby bodies.    By summing the eccentricities
of all the planets it is possible to average over the secular oscillations.  The smoothed sum of the planet eccentricities shown in the top subpanel of Figure \ref{fig:e}a shows locations where stronger jumps in eccentricity 
of the entire system occur and these correspond to times when first order resonances are affecting the system.
We can interpret the slow increases in planet eccentricity during the integration
as due to first order three-body resonances.  This follows as the zero-th order resonances should
not affect the eccentricities and there are no strong nearby two-body resonances.
As the system wanders in semi-major axis first order three-body resonances are crossed leading 
on average to the slow eccentricity evolution evident in Figure \ref{fig:e}a.

\section{Resonance Overlap Criterion}

We consider whether the number density and widths of three-body resonances are sufficient that
they are likely to overlap.  We estimate the number density of three body resonances
and multiply this by the width of the resonances.  The result is a filling factor that if greater than 1 implies
that the three-body resonances overlap and so can induce chaotic behavior in the system.  
The `resonance overlap criterion' for the onset of chaotic behavior was pioneered
by Chirikov in 1959 (see \citealt{chirikov59,chirikov79,lichtenberg92}).
A similar approach has been used to estimate the width
of the chaotic zone near a planet's corotation resonance \citep{wisdom80,murray97,quillen06a} and the
onset of chaotic behavior in other settings \citep{chirikov79,holman96,mudryk06,mardling08}.
Three-body  resonance overlap has been seen in numerical
studies of the outer Solar system  \citep{guzzo05}.

Each zero-th order
three body resonance is specified by two integers $p,q$ and three consecutive planets.
Given a particular $p$ value we can first consider the separation between the three-body resonances with 
different $q$ values.
We consider an equally spaced system with separation set by $\delta$.  Let $y$ be the ratio
of mean motions between consecutive planets $y = (1+\delta)^{-3/2}$.   The vector 
of mean motions for three consecutive planets ${\bf n} = (1, y, y^2)$ and 
distance to resonance $B = {\bf n} \cdot {\bf z}$ (equations \ref{eqn:AB}, \ref{eqn:B}).
This gives  $B = p - (p+q)y + qy^2$.  Solving for $y$ when $B=0$ we find that on resonance $y=p/q$.   
We can differentiate this with respect to $q$ to find the distance between resonances.
Given a value for $p$, the distance between resonances with adjacent $q$ values 
(in terms of differences in mean motions or
in terms of differences in $\delta$) is of order $ d\delta = p/q^2$.    
For small $\delta$, $y$ is near 1 and so on resonance $p \sim q$.    
Consequently the number density of three body resonances with $p$ is (and estimated
from the separation $d\delta$) is
\begin{equation}
\rho_\delta(p)  \sim p.
\end{equation}
The number density is approximately in units of the mean motion of the innermost body involved in the resonance.

We now consider the width of each resonance.
Consider $\delta = \delta_{pq} + x$ where 
 $\delta_{pq}$ is on resonance with ${\bf z} \cdot {\bf n}(\delta_{pq}) = 0$.
We estimate the distance to resonance  (equations \ref{eqn:AB}, \ref{eqn:B})
\begin{equation}
B \approx {\bf n}(\delta_{pq} +x) \cdot {\bf z} \sim {3 \over 2} (p-q) x
\end{equation}
To be near or within the resonance
(determined by the condition given in equation \ref{eqn:inres} or distance
to resonance is smaller than $\sqrt{2}$ times the libration frequency) 
\begin{equation}
{3 \over 2} |(p-q)x| \lesssim 6 m p \delta_{pq}^{-1} |\ln \delta_{pq}| \exp(- \delta_{pq} p), 
\end{equation}
or 
\begin{displaymath}
|x| \lesssim {4 m \over |p-q|} \delta_{pq}^{-1} |\ln \delta_{pq}| \exp(- \delta_{pq} p).
\end{displaymath}
Here we have  used  equation (\ref{eqn:om}) for the libration frequency.
Multiplying by a factor of two so as to cover both sizes of resonance and assuming $|p -q| \sim 1$ appropriate
for resonances when $\delta$ is small, the resonant width (corresponding to a range in $\delta$ or $2 |x|$) is
\begin{equation}
w_\delta(p) \sim 8 m p \delta_{pq}^{-1}  |\ln \delta_{pq}| \exp(- \delta_{pq} p).
\label{eqn:wd} 
\end{equation}

We combine the resonant width with the number density to estimate a three-body resonance filling factor.
For each $p$ the number density of resonances  times their width is 
\begin{equation}
\rho_\delta(p) w_\delta(p) \sim 8 m p^2  \delta^{-1}  |\ln \delta| \exp(- \delta p).
\end{equation}
To estimate the total filling fraction, $f_3$, of three-body resonances we integrate the previous expression 
over all possible $p$ values;
\begin{eqnarray}
f_3 &\sim& \int_{p=1}^\infty \rho_\delta(p) w_\delta dp \nonumber \\
 &\sim & 8 m \delta^{-4} |\ln \delta|.
 \label{eqn:f3}
\end{eqnarray}
When $f_3 \gtrsim 1$ the zero-th order three-body resonances are sufficiently numerous and 
wide that they are likely to overlap.

For the simulation shown in Figure \ref{fig:threejump} we compute $f_3 \sim 0.75$ placing the system 
near the regime of
resonance overlap.   This is perhaps not surprising as we found that the system was influenced by both
the ${\bf z} = (6,-13,7)$ and $(5,-11,6)$ resonances.
Inverting the above equation with $f_3=1$ we find that  resonance overlap occurs when
\begin{equation}
\delta \lesssim 2 m^{1/4}
\label{eqn:mquarter}
\end{equation}
The $m^{1/4}$ form of the criterion may be related to the slopes for stability or crossing timescales measured by 
\citet{chambers96}.

Were we to take into account the possible combinations of consecutive planets (e.g, for $N=5$ there
are three groups of consecutive planets) and first order
resonances, the modified filling factor would be somewhat larger than computed in equation (\ref{eqn:f3}).
We expect the first order resonances to initially be weaker than the zero-th order resonances but because
of the additional possible angle combinations there there are few times more of them.  Consequently
first order resonances may contribute to the overlap criterion.
A more accurate criterion would likely cover the regime integrated by numerical studies where the separation
ranges from 0.5 -- 4  $\times m^{1/4}$ (e.g., \citealt{chambers96}).  The exponential dependence 
in the Laplace coefficient on interplanetary distance (equation \ref{eqn:lapapprox}) 
implies that three-body resonances
for non-consecutive combinations of planets are unlikely to be strong.  This explains why the
crossing or stability timescale for equidistant closely spaced systems is 
only weakly dependent on the number of planets when the number $N \gtrsim 5$ \citep{chambers96}.
The resonance overlap criterion suggests that there is a critical separation value that separates
two regimes, an inner one at small $\delta$ governed by instability from overlapping three-body resonances
and an outer one that is much more stable.  Two separate regimes and a transition
from one to another at larger separations does seem to be exhibited in
numerical integrations \citep{smith09}.
As there are fewer combinations of consecutive planets for low $N$ we would expect this transition 
would occur at smaller separations; this too is seen in numerical integrations (Figure 13 by \citealt{smith09}). 

\section{Crude Estimates for Diffusion}

We expect that the resonances that overlap would often be similar in size, as illustrated
in the integration shown in Figure \ref{fig:threejump} where the $p=5,q=6$ and $p=6,q=7$
three-body resonances were both important.  Assuming full  resonance overlap,
we estimate a diffusion coefficient for wander in the Poincar\'e coordinate  
\begin{equation}
D_\Lambda \sim (\Delta \Lambda)^2 \omega_{pq} \sim 8 p^2 \epsilon_{pq}^{3/2} A^{-1/2}
\end{equation}
where we use equation (\ref{eqn:Lai}) for the changes in $\Lambda$ and have assumed
that these changes take place on a timescale equal to the libration frequency  (equation
\ref{eqn:omega}).  This type of estimate is similar to those explored by \citet{chirikov79}.
Using approximations for these quantities (equations \ref{eqn:om} and \ref{eqn:AB})
%and setting $p \sim \delta^{-1}$ 
we estimate
\begin{displaymath}
D_\Lambda \sim 2 p m^5 \delta^{-3} |\ln \delta|^3.
\end{displaymath}
The diffusion coefficient is largest for the highest $p$ value which has $p \sim \delta^{-1}$ because
the Laplace coefficients drop exponential at higher p.  Setting $p \sim \delta^{-1}$
\begin{equation}
D_\Lambda \sim  2 m^5 \delta^{-4} |\ln \delta|^3
\label{eqn:DL}
\end{equation}
corresponding to a diffusion coefficient in semi-major axis of
\begin{equation}
D_a \sim 8 m^3 \delta^{-4} |\ln \delta|^3.
\label{eqn:Da}
\end{equation}

An upper limit for a crossing timescale would be the time it takes for the semi-major axis to wander a distance 
approximate equal to the interplanetary spacing 
$\delta$ or 
\begin{equation}
t_{u} \sim \delta^2/D_a.\end{equation} 
Using our estimate for the diffusion coefficient this gives
\begin{equation}
t_{u} \sim {1 \over 8} m^{-3} \delta^6 |\ln \delta|^{-3}.
\end{equation}
For the system we show in Figures \ref{fig:threejump} and \ref{fig:e} this corresponds to
$t_{u} \sim 3 \times 10^7$ years, and about 3 orders of magnitude greater than expected from the
fit to the crossing timescales \citep{faber07} or inferred from Figure 3 by \citet{chambers96}.  
Consequently this timescale severely overestimates the crossing timescales.
  
It is likely the eccentricity evolution must be considered as even small increases in eccentricity can strongly
affect the stability or crossing timescales \citep{zhou07}.  There are many first order three-body resonances
but eccentricity increases due to them are very small.  There are fewer two-body first order resonances
but eccentricity increases due to them could be high if the system wanders into one.
The minimum eccentricity of a body in the vicinity of a first order mean motion resonance scales
with $m^{1/3}$ (e.g.,  \citealt{quillen06}, table 1, or \citealt{mustill11}).  This is a small power, and so not small for the regime covered by numerical integrations that have interplanetary separations of order a few to a dozen
Hill radii.
We adopt the ansatz that the crossing timescale is set by the time it takes for the system to
wander into a first order mean motion resonance among two consecutive bodies.

We first estimate the distance that one planet must wander (due to 
the three-body resonances) before it encounters a first order two-body mean motion resonance. 
In the limit of high $q$, first order resonances (with mean motions with a ratio of $q:q-1$) 
are separated in semi-major axis by 
$\Delta a \sim q^{-2}$.  For an interplanetary separation of $\delta$
the nearest first order two-body resonance likely has $q \sim \delta^{-1}$.  
Thus the distance between resonances is $\Delta a \sim \delta^{-2}$.

\begin{figure}
\includegraphics[width=8cm, trim=0mm 5mm 0mm 5mm]{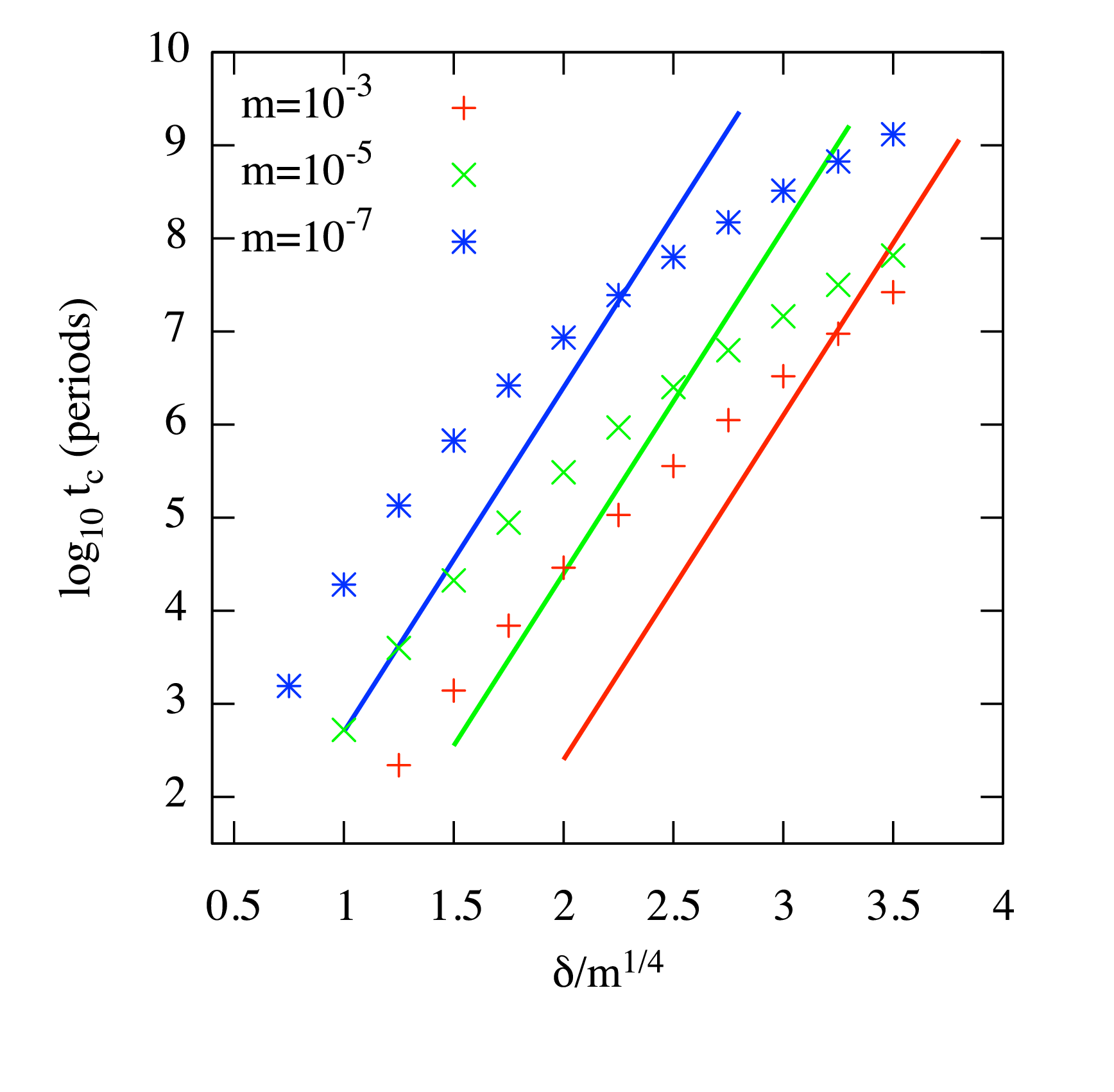}
\caption{A comparison of numerical to estimated crossing timescales.  The crossing time estimated from the time it takes the system to cross a first order mean motion resonance between two consecutive bodies (calculated with equation \ref{eqn:t2}) is shown as points for three
different planet mass ratios.  The function fit
to numerically measured crossing times 
(using equation 2 by \citealt{faber07}) is shown as line segments for the same
three planet mass ratios.
}
\label{fig:diff}
\end{figure}

Using our above estimated diffusion coefficient for semi-major axis wander (equation \ref{eqn:Da}),
the time it takes to cross a first order two-body mean motion resonance due to wander in semi-major axis 
would be of order
\begin{equation} 
t_{2} \sim (\Delta a)^2/D_a \sim \delta^{-4}/D_a \sim {1 \over 8} m^{-3} \delta^8 |\ln \delta|^{-3}.
\label{eqn:t2}
\end{equation}
In Figure \ref{fig:diff} we show a comparison between crossing times predicted with the above
$t_2$ timescale and crossing times measured numerically.  The numerically measured
timescales are shown with the fit to the numerically measured value by \citet{faber07}. 
This estimate is within two orders of magnitude of the relation found
numerically and overestimates the crossing time at small separations. Nevertheless it is the
first analytical derived estimate of crossing time that covers an appropriate range in parameter space. 
We note that because of the high power of $\delta$ in the above equation our power law relation
is nearly as steep as the numerically measured times that have primarily been fit with exponential functions.
On Figure \ref{fig:diff} the resonance overlap boundary would be at a constant value  of $\delta/ \mu^{1/4}$
corresponding to a vertical line on the right hand side of this plot.  Numerical integrations cover the range
of $\delta/\mu^{1/4}$
shown in this plot hence we expect that the overlap criterion line should lie on the right hand side 
of the plot with $\delta/ \mu^{1/4} \sim 3.5$.  
Our predicted location (equation \ref{eqn:mquarter}) for the onset of
three-body resonance overlap is about a factor of two or so too low as this line lies in the middle rather
than the right hand side of the plot.

The above crude estimates for a diffusion coefficient (equation \ref{eqn:DL}) and 
a crossing time (equation \ref{eqn:t2})
are strong power law functions of both mass and interplanetary separation.
Their strong dependence on separation suggests that the exponential forms fit to the numerically
measured stability or crossing
timescales \citep{chambers96,duncan97,zhou07,chatter08,smith09} 
might in future be accounted for through chaotic motions induced
by three-body resonances.   Perhaps the exponential form has provided a good fit because of the limited range
of times over which these systems can be integrated and because the diffusion rate is such a strong function of
mass and separation.  If three-body resonances overlap then there is no underlying mathematical reason
(related to Arnold diffusion or the Nekhoroshev theorem) 
that would predict an exponential dependence on interplanetary separation and mass.
We suspect that
only outside the regime of three-body resonance overlap could a true long timescale exponential dependence on
planet mass and separation be recovered.

\section{Summary and Discussion}

In this paper we have considered the role of three-body resonances for an idealized equal mass, 
uniformly spaced, but closely packed initially low eccentricity co-planar multiple planet system.
We have estimated the strengths and libration timescales for zero-th order (in eccentricity) 
three-body resonances using an asymptotic approximation to the $s=1/2$ Laplace coefficient.    
Two conserved quantities
are present relating variations in semi-major axes between the bodies affected by the resonance.
These variations and the Laplace angles are useful for identifying the effect
of three-body resonances in numerical integrations of multiple planet systems.

By estimating the number and widths of the three-body resonances,
we have derived an approximate resonance overlap criterion.  We find that zero-th order three-body resonances are likely to overlap
when the separation between planets $\delta \lesssim 2 m^{1/4}$.  
The resonance overlap criterion is close to the regime covered by numerical integrations of multiple planet systems
exhibiting instability \citep{chambers96,duncan97,zhou07,chatter08,smith09}, 
suggesting that the instability seen in  
integrations of closely spaced multiple planetary systems is due to chaotic behavior associated with
multiple three-body resonances.   We note that previous studies of asteroids have also attributed  
chaotic behavior to three-body resonances \citep{murray98,nesvorny98}.

Our resonance overlap criterion lies near but within the regime covered by numerical integrations 
exhibiting instability implying that we have underestimated the filling factor of 
resonances by a factor of a few.  
However, we have not 
taken into account first order three-body resonances, the different combinations of consecutive planets,
indirect terms in the Hamiltonian
and we have only crudely estimated resonance strengths.  Future works can improve upon the overlap criterion by
expanding and improving the calculation.  

For spacings  larger than the overlap criterion boundary
three-body resonances should not be as dense and the probability of resonance overlap drops.   
We postulate that there is a region of long
timescale stability at large separations.  This region and the transition between the two regimes 
is likely the reason for measured changes in slope of crossing time
versus separation and a strong increase in crossing time at large separations 
(see Figures 1-4 by \citealt{smith09}).    The filling factor of three-body resonances
should be lower for systems with fewer planets because there are fewer combinations of consecutive
planets and so fewer strong three-body resonances.    Thus we expect the transition to a more stable regime
would occur at smaller planetary separations when there are fewer planets.  This also
is seen in numerical integrations  (see red points in Figure 13 by \citealt{smith09}).

We have attempted to predict diffusion rates using three-body resonances.     The timescale
to wander a distance of the interplanetary separation grossly overestimates the crossing timescale
whereas that to diffuse until the system crosses a first order mean motion resonance between
two bodies overestimates the crossing timescale by 1 or 2 orders of magnitude at small separations.
This estimated timescale depends on the 8-th power of the interplanetary
spacing suggesting that exponential functions have primarily been successful at fitting
numerically measured crossing timescale because of the strong dependence on separation of the three-body resonances.
Future work could strive to improve upon these gross estimates.
Crossing timescales are not directly related to diffusion coefficients and the dynamics may
be intermittent and diffusion anisotropic (e.g., \citealt{shev10,guzzo05}).
To better account for or predict the crossing timescales perhaps 
Lyapunov timescales and diffusion coefficients (i.e., eccentricity growth rates and  rates of
wander in semi-major axis) could be measured directly from numerical integrations.
These then may be easier to understand with 
analytical estimates such as explored here. 

Here we have considered equal mass, equidistant coplanar systems.  However much of 
the framework developed here could be applied to less ideal systems such
as closely spaced multiple planet extrasolar planetary systems.   We remind the reader
that here we have focused on
systems that are in three-body resonances but are not in strong two-body resonances.
Three body resonances may also be important in these systems but 
calculations are likely to be more challenging in this setting.
%applications     
Diffusion in semi-major axis seen in simulations of closely spaced satellite systems, e.g., the Uranian
satellite system, \citep{duncan97,showalter06,dawson10} and the Kepler 11 system \citep{lissauer10}
might in future be interpreted in terms of variations arising from three-body resonances.

\vskip 0.3 truein

Acknowledgements.    This work was in part supported by NSF through award AST-0907841.   
We thank the Isaac Newton Mathematical Institute for hospitality and support during the fall of 2009 where 
this work was begun.

This work could not have been carried out without helpful discussions with 
Pierre Lochak, Ivan Shevchenko, 
J.-L. Zhou, Eric Ford, Adam Lanman, and Rob French.

I thank the referee for a thorough and careful reading of this manuscript leading to many corrections
in calculation.

\end{document}